# The Impact of Virtual Mirroring on Customer Satisfaction

Gloor, P. A., Fronzetti Colladon, A., Giacomelli, G., Saran, T., Grippa, F.

# The Impact of Virtual Mirroring on Customer Satisfaction

Gloor, P. A., Fronzetti Colladon, A., Giacomelli, G., Saran, T., & Grippa, F.

**Abstract**

We investigate the impact of a novel method called "virtual mirroring" to promote self-reflection and impact customer satisfaction. The method is based on measuring communication patterns, through social network and semantic analysis, and mirroring them back to the individual. Our goal is to demonstrate that self-reflection can trigger a change in communication behaviors. We illustrate and test our approach analyzing e-mails of a large global services company by comparing changes in customer satisfaction associated with team leaders exposed to virtual mirroring (the experimental group). We find an increase in customer satisfaction in the experimental group and a decrease in the control group (team leaders not involved in the virtual mirroring process). With regard to the individual communication indicators, we find that customer satisfaction is higher when employees are more responsive, use a simpler language, are embedded in less centralized communication networks, and show more stable leadership patterns.

## 1. Introduction

Just like the human body shivers when it has a fever, an organization frequently has a vague feeling that something is amiss, but is unable to pinpoint what is wrong. Similarly to the thermometer measuring the health of the body, we propose a novel approach to assess organizational health by calculating a series of communication metrics between individuals in the organization. Our goal is to demonstrate that offering individuals the opportunity to reflect on their own communication behaviors has the potential to change those behaviors and ultimately affect customer satisfaction.

Since the effects of feedback interventions - or mirroring sessions as we call them - on performance is far from being clearly explained (Hattie & Timperley, 2007; Kluger & DeNisi, 1996), our goal is to offer empirical evidence that there might be a positive association between increasing awareness of your own communication behaviors and an improvement of such behaviors. Just like mirror neurons put the self and the other back together and map the actions of the other into the self (Iacoboni, 2009 p. 155), our virtual mirroring process is grounded on the idea that self-awareness requires socialization and continuous dialog on the impact that our words and behaviors have on others.

This paper describes the results of a two-year experiment that started in June 2012, where we assessed improvements in customer satisfaction using a virtual mirroring process that allows employees to learn about their own communication behavior tracked through e-mail analysis. In this project we involved leaders of 26 large accounts in monthly virtual mirroring sessions, where the communication characteristics of the teams working with clients were shared and discussed in plenary and individual sessions. To this purpose, we measured the structure of the communication network, looking at who is interacting with whom, the average complexity of the vocabulary used, as well as the responsiveness of employees to customers' emails. In this study, a social network structure is defined as the structure resulting from the regularities in the patterning of relationships among organizational members (Battistoni & Fronzetti Colladon, 2014; Wasserman & Faust 1994).

## 2. Virtual Mirroring as a way to promote Self-Awareness and Organizational Communication

Brainstorming or learning sessions represent a powerful methodology to educate members on their personal communication and learning styles. The process of looking at ourselves 'in the mirror' is important to create self-awareness, foster generative learning and shape team learning (Kluger & DeNisi 1996). Recent studies illustrated the importance of conducting learning sessions by presenting results of social network analysis to organizational leaders with the potential to trigger a self-organizing feedback loop (Grippa, Gloor, Bucuvalas, & Palazzolo, 2012). These sessions allow members to reflect on their own communication styles, to look at their own areas of improvements, and to perceive themselves under a new perspective.

Mirroring and feedback sessions are commonly used in industrial and social psychology studies (Pritchard et al., 1988), and are widely adopted by consultants as a way to promote organizational change (Ramos, 2007). Virtual mirroring may encourage individuals to change behaviors in pursuit of targeted outcomes and might help identify opportunities for leaders to support innovation. For example, Gesell et al. (2013) conducted a social network study where group leaders would receive a network map and specific data-driven recommendations on how to increase group connectivity. If the network was cohesive at session four, then the leaders would be instructed not to alter their teaching methods. Their sessions empirically guided program activities and resulted in increased group cohesion.

In virtual mirroring sessions, social network diagrams are presented and described to participants, along with individualized reports. These reports are widely used as an effective tool to help team members self-diagnose communication patterns and promote a process of behavioral change. After using feedback interventions with communities of practice, Cross and colleagues noted how "*Often one of the most effective interventions is simply to ask people to spend five minutes, either on their own or in groups of two or three, to identify what they 'see' in the map and the performance implications for the group*" (Cross et al., 2002, p.11). Measuring individuals' communication patterns is the first step to create and nurture a climate of reciprocity with regard to information exchange and collective learning. Promoting awareness can be realized via a process of self-reflection, or mirroring, which draws attention to important aspects of organizational development and individual behavioral patterns. Self-awareness is defined as the individual ability to introspect and recognize oneself as an individual separate from the environment and other individuals (Ramachandran, 2011). Collective awareness requires team members to create and nurture a climate of reciprocity with regard to information exchange. In this context, interaction plays a key role in the creation and maintenance of collective awareness and in the development of a shared awareness about what they are engaged in (Weisband, 2002).

In a multi-experiment study, Wicklund and Duval (1971) used a mirror to induce self-awareness in the participants and found that performance was better for the groups that used a mirror during the study. In support of the Hawthorne effect (Gillespie, 1991), this stream of research seems to suggest that environmental factors - such as mirrors, an audience, being observed or recorded - have the potential to induce self-awareness and improve performance.

The ability to reflect on your sense of self is an important component of self-awareness. Since self-awareness might not be sufficient to trigger the desired change when an individual lacks fundamental skills, it is important to encourage readiness by offering the necessary tools to interpret the results of the mirroring sessions and set personal goals (Zimmerman, 2002).

Recently, Pentland and other researchers (Pentland, 2008, 2012; Gloor et al., 2007) studied how specific communication patterns account for differences of team performance and determined that communication exposure determines behavior and performance: "*People tend to learn more by copying others and by "trying things out" versus working in isolation. Because of this, the more communication a person is exposed to, the more information he or she can harvest and put to use*" (Pentland, 2012, p.3). Gloor (2005) developed a methodology called Knowledge Flow Optimization (KFO) that uses social network analysis to monitor communication among team members, comparing it with performance, and mirroring back results to participants. The KFO methodology, which comprises four main steps (Discover-Measure-Optimize-Mirror), has been applied to track and support the growth of organizations from project start to completion (Gloor, 2005). Optimal communication structures vary depending on the institutional context and changing its form or content can lead to very different behaviors and results (Král & Králová, 2016). For example, for call center staff and nurses in a hospital setting, more hierarchical styles lead to better results (Olguin et al., 2007), while life sciences researchers deliver better results in decentralized collaboration networks (Grippa et al., 2012). In the mirroring process, the communication patterns of individuals and teams are shown to team members, together with information about communication patterns of the most successful individuals and teams (Grippa et al., 2012). This insight is based on the Hawthorne principle (Gillespie, 1991): telling a group of people that they are being monitored and what the desirable communication patterns are will get them to change their behavior towards the desired outcome.

Based on the empirical evidences and theoretical contributions described above, there is a high likelihood that a virtual mirroring process will lead to increased self-awareness and self-reflection on communication behaviors, which creates the condition for behavioral change. Since people tend to align their behavior with their standards when they become aware of any discrepancy (Duval & Wicklund, 1972), we conclude that virtual mirroring sessions have the potential to stimulate self-evaluation and foster behavioral change.

The dependent variable used in this study is customer satisfaction, which is widely considered an antecedent of customer loyalty, retention, word of mouth and firm profitability (Anderson, Fornell, & Lehmann, 1994; Bearden & Teel, 1983; Torres & Tribó, 2011; Wirtz & Lee, 2003). If a firm has a strong customer loyalty, this typically impacts the firm's economic return. Satisfied customers tend to be more willing to pay for the benefits they receive and are more likely to accept increases of prices (Anderson et al., 1994). Most of the variables commonly reported as having an effect on customer satisfaction involve employees' behaviors and include their level of friendliness, courtesy, competence, and support. This suggests that customer satisfaction is tightly related to employee satisfaction. Excellence in service requires building skills such as listening, empathy, empowerment, and a culture fostering innovation and creativity (Gremler & Gwinner, 2000). The key to acquiring and retaining today's customers is to deeply engage them in meaningful conversations, creating a connection with the company and nurturing a two-way dialogue that creates active participation. The traditional methods to detect customer satisfaction are heavily relying on surveys, and most of the time they lack timeliness and details.

Figure 1 presents the virtual mirroring process. The independent variables are the metrics of social network and semantic analysis, organized on three dimensions: degree of connectivity, use of language, and degree of interactivity (Gloor, 2006, p. 175; Zhang, Gloor and Grippa, 2013). These indicators are the same used during the virtual mirroring sessions to illustrate organizational members their communication behaviors.

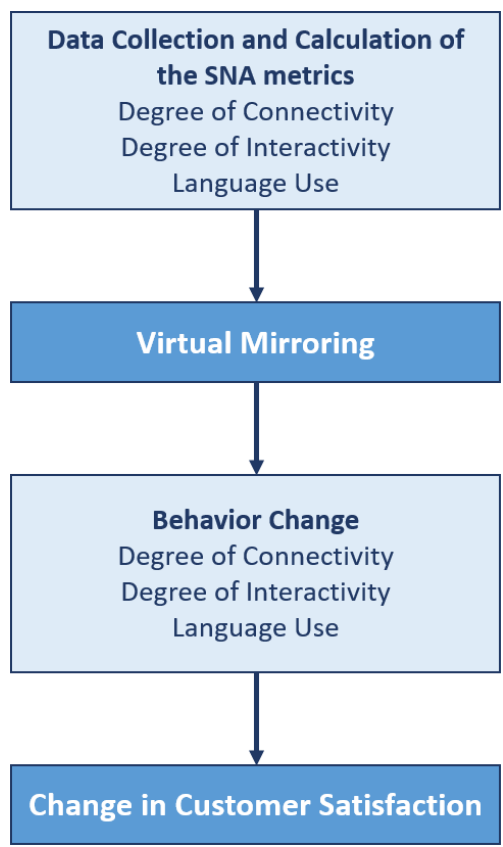

**Figure 1.** The Virtual Mirroring process.

### 2.1. Study Hypotheses

Our first hypothesis relies on the assumption that the feedback provided via virtual mirroring sessions represents a self-reflection opportunity that can create a change in the communication style.

*(H1) Exposing members of the organization to their own communication patterns - triggering their self-awareness - will increase customer satisfaction.*

Our second hypothesis is based on the assumption that customer satisfaction increases when the communication style is more direct with customers. A direct commitment to individual customers has been recognized as a crucial factor for generating enhanced performance (Brodie et al., 2011). As demonstrated by Webster & Sundaram (2009), an affiliative communication style will lead to greater customer satisfaction, especially when customers are in a relatively complex service situation, and they need a highly affiliative provider to reduce anxiety and tension. In their study on the impact of staff empowerment and communication style on customer evaluations, Sparks and colleagues (1997) found strong evidences that customers are more satisfied when they interact with employees who are fully empowered by institutional policies and procedures to make immediate decisions and take actions to resolve customers' concerns. They also found that more positive evaluations are associated with a communication style characterized by empathy and personalization. Current research demonstrates that the ability to connect on a personal level with customers, offering prompt responses to their needs, increases their loyalty and satisfaction (Gwinner, Bitner, Brown, & Kumar 2005; Roman & Iacobucci, 2010). As demonstrated by Roman & Iacobucci (2010), service providers who communicate using an adaptive, accommodative, or affiliative styles, are more likely to elicit positive reactions from customers, generating higher customer satisfaction and customer perceptions of service quality. When customers perceive that a supervisor is being called in to deal with the problem, their level of satisfaction decreases. This implies that customers are more satisfied when they believe that the service provider is the direct link in dealing with their problem. This would lead to our second hypothesis:

*(H2) Customers are more satisfied when communication is more direct.*

Our third hypothesis explores the connection between customer satisfaction and complexity of the e-mail subject lines (Broennimann, 2014). Extensive studies in the area of service marketing show that effective communication requires the formal as well as informal sharing of meaningful and timely information between a client and a provider in an empathetic manner (Brodie et al., 2011). Customers tend to be more satisfied when providers adopt an affiliative communication style aiming to reduce anxiety and address their concerns (Webster & Sundaram, 2009). Other studies demonstrate that a frequent, clear and timely communication of a service provider contributes to the perception of service quality (Morgan & Hunt, 1994). A simpler and more direct communication with client helps reach the goal of keeping clients informed in a language that they can understand. Customers' concerns should be addressed with a language that is not overly complex. As illustrated by Pennebaker (2013), our language can provide insights into our feelings; therefore, service providers should try to avoid words that could be misinterpreted and increase anxiety. Based on the extant literature, we hypothesize that:

*(H3) Customers are more satisfied when providers engage with them using email subject lines that are worded in a simpler shared language.*

Our fourth hypothesis is based on studies showing that productivity, innovation and learning are encouraged when organizational members display communication patterns that are more focused and dedicated to each single customer. While our second hypothesis reflected the "connectivity" dimension of our model (number of ties to other nodes), H4 is specifically focused on the "interactivity" dimension, and emphasizes the email response time. Prior research suggests that customer satisfaction is positively correlated with a lower response time to customers' emails (Davidow, 2003) and negatively correlated with delays (Boshoff, 1997; Sirdeshmukh, Singh & Sabol, 2002). As recently demonstrated by Kooti et al. (2005), the volume of incoming email affects behaviors of recipients and the length of time it takes them to reply to emails. When employees have the opportunity to focus on fewer projects and their attention is not spread across too many tasks, they will have more time and energy to commit on a deeper level, responding faster and increasing customer satisfaction (Merten & Gloor, 2010). A fast response time reflects a positive interpersonal adaptive behavior defined by Bock, Mangus, & Folse (2016) as the customization of the interpersonal elements of the service process (including communication and presentation style) within the customer-employee interaction. We therefore hypothesize that:

*(H4) Customers are more satisfied interacting with providers who respond more rapidly to their emails.*

Our fifth hypothesis relates to the shift in betweenness centrality, which we define as "oscillation". This metric helps identify employees who move from highly central positions to more peripheral positions in the communication network. As suggested by Davis and Eisenhardt (2011), to reduce the costs of alternating control and decision making in situations where an immediate action is required, it could be appropriate to adopt a more "dominating" leadership style. Since service providers are usually asked to be reliable and predictable, we speculate that customers are more satisfied in stable network structures that change little. Customers would like to interact with the providers on a fixed schedule, leading to as little oscillation as possible. As suggested by Sparks and colleagues (1997) customers are more satisfied when they interact with service providers who are empowered to take immediate actions to solve customers' problems. Since role ambiguity tend to be negatively correlated with employees' ability to go the extra mile to help customers (Malhotra & Ackfeldt, 2016), we expect customer satisfaction to increase when roles are clearly defined and stable. Our final hypothesis states that:

*(H5) Customer satisfaction increases when betweenness centrality oscillation decreases.*

## 3. Research Design

In order to measure network structure and responsiveness to customers, we combined metrics such as the number of emails exchanged between clients and employees and their average response time (Gloor, Almozlino, Inbar, & Provost, 2014). To calculate these metrics we applied methods and tools of dynamic social network analysis (Allen, Gloor, Fronzetti Colladon, Woerner, & Raz, 2016; Carley et al., 2009; Fronzetti Colladon & Remondi, 2017; Valente, 2012; Wasserman & Faust, 1994) in conjunction with more recent developments of semantic analysis (Grant, 2013; Losada, 1999; Pennebaker 2013; Pang & Lee, 2008; Whitelaw et al., 2005). The degree of connectivity in our model measures how well connected individuals are using social network metrics, such as actor betweenness centrality, group betweenness centrality and group degree centrality (Wasserman & Faust, 1994). In general, we can consider the email network as an oriented graph made of a set of n nodes (email accounts) – referred as G = {$g_1$, $g_2$, $g_3$ … $g_n$} – and of a set of m oriented arcs (emails) connecting these nodes – referred as A = {$a_1$, $a_2$, $a_3$ … $a_m$}. This graph can be represented by a sociomatrix X made of n rows and columns, where the element $x_{ij}$ positioned at the row i and column j is bigger than 0 if, and only if, there is an arc ($a_{ij}$) originating from the node $g_i$ and terminating at the node $g_j$. When the elements of X are bigger than zero, they represent the weight of the arcs in the graph ($x_{ij}$). If arcs are not weighted, we consider a dichotomized sociomatrix where each element can only assume the value 0 or 1.

Group Degree centrality is a measure expressing how centralized a network is and if it is dominated by few very well connected actors or else all degrees are similar (Wasserman & Faust, 1994). A lower GDC indicates that employees are communicating with customers in a more direct way, rather than spreading the energy on multiple customers at the same time. Starting from the formula used to calculate the degree centrality for the node $g_i$:

$$D_C(g_i) = \sum_{j=1}^{n}(x_{ij} + x_{ji})$$

We can define the group degree centrality as illustrated by Wasserman and Faust (1994), considering $D_C(g^*)$ as the largest observed centrality value in the group:

$$GD_C = \frac{\sum_{i=1}^{n}[D_C(g^*) - D_C(g_i)]}{[(n-1)(n-2)]}$$

Individual betweenness centrality scores, indicate how frequently a node $g_i$ is in-between the shortest paths that connect every other pair of nodes:

$$B_C(g_i) = \sum_{j<k} \frac{d_{jk}(g_i)}{d_{jk}}$$

Where $d_{jk}$ is the number of shortest paths linking the generic pair of nodes $g_j$ and $g_k$, and $d_{jk}(g_i)$ is the number of paths which contain the node $g_i$. This measure can be standardized dividing it by $[(n-1)(n-2)/2]$, which is the total number of pairs of actors not including $g_i$. Starting from this formula, we refer to Wasserman and Faust (1994) to calculate the group betweenness centrality score, which expresses the heterogeneity of the betweenness for the members in a group:

$$GB_C = \frac{2\sum_{i=1}^{n}[B_C(g^*) - B_C(g_i)]}{[(n-1)^2(n-2)]}$$

Where $B_C(g^*)$ is the largest $B_C(g_i)$ score for the actors in the group.

Lastly, we operationalize the measure of the group betweenness centrality oscillations (Kidane & Gloor, 2007) counting the local maxima and minima of function $f(t)=B_C(g_i, t)$ within time interval $[t1,t2]$. There is a local maximum for time t at point t*, if there exists some $\varepsilon > 0$ such that $f(t^*) \geq f(t)$ when $|t - t^*| < \varepsilon$. Similarly, we count the local minima at t*, if $f(t^*) \leq f(t)$ when $|t - t^*| < \varepsilon$. Betweenness centrality oscillations for actor $g_i$, over the time window ws, are therefore given by the formula:

$$B_C Osc(g_i) = \#local\ minima(g_i)^{ws} + \#local\ maxima(g_i)^{ws}$$

The group betweenness oscillation score is obtained summing up all the oscillations found for the individual actors in the group.

The second dimension of our analysis reflects the complexity of the language used, calculated as the average difficulty of the vocabulary in each email subject line. Complexity is calculated as the likelihood distribution of words within an e-mail text, i.e. the probability of each word of a dictionary to appear in the text (Brönnimann, 2014). The algorithm used to calculate complexity uses the well-known term frequency/inverse document frequency information retrieval metric (Brönnimann, 2014). It measures how popular a word is in an individual message compared to the word's occurrence in the overall text collection; a message that uses more comparatively rare words has a higher complexity. The third dimension, interactivity, is measured considering the average response time (ART) of an actor to respond to a message (Gloor et al., 2014).

We used the Net Promoter Score (NPS) to operationalize our dependent variable, customer satisfaction (Reichheld, 2003). NPS is based on a simple question: "How likely is it that you would recommend our company to a friend or colleague?" on a scale from 0 to 10, where 10 means "extremely likely" and 0 means "not at all likely". NPS is then computed by subtracting the percentage of "detractors" (scores 0-6) from the percentage of "promoters" (scores 9-10). As noted by Reichheld: "*Tracking net promoters - the percentage of customers who are promoters of a brand or company minus the percentage who are detractors - offers organizations a powerful way to measure and manage customer loyalty*" (Reichheld, 2003, p. 7).

### 3.1. Experimental Setup

The design of this experiment involved scheduling virtual mirroring sessions on a monthly basis. Each virtual mirroring session allowed participants to review their communication patterns, so as to reflect on their communication behaviors and discuss insights with colleagues (Shadish, Cook, & Campbell, 2002). In these monthly review sessions their communication variables were shown by a company social network analyst to the account leaders as a simple score card, normalized from 0 to 1, together with their social network picture (Gloor & Giacomelli, 2014). A review session would last from 30 minutes to one hour. A total of 26 key account leaders, whose entire e-mail with customers was collected, were exposed to monthly virtual mirroring over seven months (November 2013 to May 2014). Teams working on a single account ranged in size from a few dozen to hundreds of members. As a control group, we took 150 accounts whose e-mail could not be collected, since the employees had to use the e-mail management system of the clients. This selection criterion for the control group is exogenous to the virtual mirroring process. Both virtual mirrored accounts and control group represent mostly global Fortune 500 firms operating in similar industries, with approximately the same representation of industries in each group (finance, pharmaceutical, high-tech, retail). They also have comparable numbers of provider associates dedicated to working on the account, ranging from 10 to 500 persons on the account. We also checked for systematic differences in subjects' characteristics between experimental and control groups, which should reduce the selection threat to internal validity (Shadish et al., 2002). To this purpose we tested for differences in mean values taking into account the industry sector and found no significance. The two groups received services that are similar and comparable. While considering the mean NPS scores of October/November 2013, we noticed that the control group started on average with higher NPS scores. Future researchers that wish to replicate this study are encouraged to adopt a new random selection criterion for the experimental and control group.

For this analysis, we pulled the data from the aggregated email archives four times over a period of two years, corresponding to the time in the year when NPS was collected by the company. At each data collection point, we gathered data for two months: June/July 2012, October/November 2012, October/November 2013, and June/May 2014. This allowed us to set up our experiment on a unique dataset, which includes a large amount of information exchanged and interactions between the clients and the global services company (more than 4,500,000 emails).

We were not able to analyze the body of the email due to privacy issues, thus our content analysis is based on the subject line of the email. In earlier work, we had shown that sentiment calculated from the e-mail body is correlated to team performance variables (Gloor & Paasivaara, 2013). While sentiment and complexity in language of message body and subject line are only weakly correlated. We tested this approach using another e-mail corpus and compared sentiment of subject line and message body of the 725 most active actors of the publicly available Enron e-mail archive[1]. Those results, also confirmed in (Gloor, 2016), show a significant correlation of 0.25 between sentiment of message subject and message body, and significant positive correlation of 0.10 between complexity of subject line and message content. This implies that both message body and subject line might have predictive power for team performance.

## 4. Results

Our results indicate that virtual mirroring could be an effective process to improve customer satisfaction. Measuring the change in NPS from October/November 2013 to May/June 2014 (teams were exposed to monthly virtual mirroring sessions, starting November 2013 and ending May 2014), we observed an average improvement of 5% in NPS for the 26 accounts in the experimental group exposed to virtual mirroring, while the 150 accounts in the control group showed a significant average decrease in NPS by 12% (Table 1).

---

[1] https://www.cs.cmu.edu/~./enron/

| Group | N | NPS Oct/Nov 2013 | NPS May/Jun 2014 | Mean NPS change | Standard Deviation | Levene's Test (F) | T-test (t) | Mean Difference |
|---|---|---|---|---|---|---|---|---|
| **Experimental group** | 26 | .483 | .532 | .050 | .267 | 1.948 | 2.247* | .166 |
| **Control group** | 150 | .801 | .684 | –.117 | .331 | | | |

Notes. *p<0.05.

**Table 1**. T-test for change in NPS from Oct/Nov 2013 to May/June 2014 of experimental (virtual mirrored) and control group (non-virtual mirrored).

As table 1 shows, there was a difference in the average NPS scores of the two groups at the beginning of the virtual mirroring process (October 2013). It is possible that some of the groups with a lower NPS score were more motivated to improve the satisfaction of their clients. Therefore, to further validate our findings, we suggest replicating our experiment using a control group with an initial average NPS more similar to the experimental group.

Figure 2 shows the change in monitored behaviors at the end of the mirroring sessions: employees' average response time declined, the overall language used in communications was simpler and betweenness centrality oscillations went down to indicate a shift towards more stable interactions.

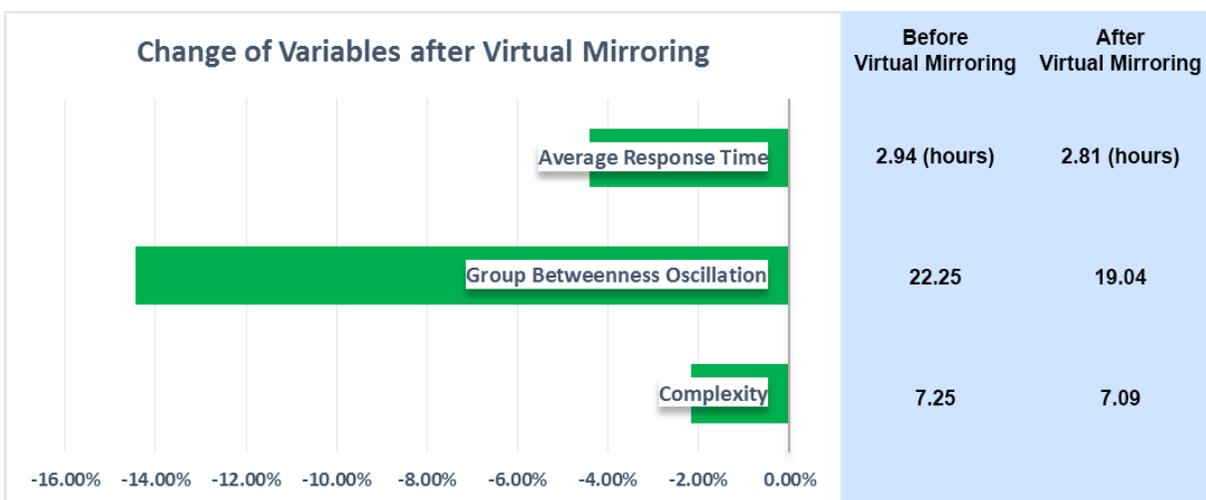

**Figure 2.** Change of variables after Virtual Mirroring (mean values).

These findings indicate that allowing employees to observe and reflect every month upon their communication behaviors is a possible explanation of the modification of their communication style, which in turn leads to more satisfied customers. As part of the virtual mirroring process, employees were invited to learn about their relative position in the communication network observing the evolution over time of metrics such as group degree centrality, group oscillation in betweenness centrality, average complexity, and average response time. As Table 2 illustrates, these interaction metrics correlate significantly with NPS values.

|   | **Variable** | **1** | **2** | **3** | **4** | **5** |
|---|---|---|---|---|---|---|
| **1** | **Customer Satisfaction (NPS)** | 1.00 | | | | |
| **2** | **Group Degree Centrality** | -.278* | 1.00 | | | |
| **3** | **Group Betweenness Oscillation** | -.352** | .185 | 1.00 | | |
| **4** | **Language Complexity** | -.257* | .024 | -.054 | 1.00 | |
| **5** | **Average Response Time** | -.289* | .217 | .124 | .101 | 1.00 |

Notes. *p<0.05; **p<0.01.

**Table 2.** Significant correlations of variables and NPS.

These correlations suggest that faster responses and simpler language positively associate with customer satisfaction; in addition, lower group degree centrality and group betweenness centrality oscillation can contribute to an improved relationship with customers. Our results show that customers are more satisfied when there is a stable point of reference in their interaction with the company, without excessive oscillations and rotations. As suggested by Davis and Eisenhardt (2011), given the costs of alternating control and decision making that are typical of rotating leadership, in certain situations "dominating" or "consensus" leadership styles may be well suited to collaborations. Since rotating leadership usually involves fluid membership, unplanned alternations, unexpected changes to the organizational direction, personal relationships that had been established with customers could be disrupted.

To further explore the predictive capabilities of our approach, we tested a series of multilevel regression models presented in Table 3.

| Variable | Model 1 | Model 2 | Model 3 | Model 4 | Model 5 | Model 6 | Model 7 |
|---|---|---|---|---|---|---|---|
| Constant | .4703** | .4654** | .4707** | .4658** | .4695** | .4171** | .4611** |
| Complexity |  | -.1976* |  |  |  |  | -.2014* |
| Group Betweenness Oscillation |  |  | -.0422* |  |  |  | -.0431* |
| Average Response Time |  |  |  | -.0487* |  |  | -.0454* |
| Group Degree Centrality |  |  |  |  | -.0404 |  |  |
| Time |  |  |  |  |  | .01855 |  |
| Variance Level 2 | .0453 | .0448 | .0370 | .0385 | .0401 | .0485 | .0319 |
| Variance Level 1 | .0473 | .0437 | .0469 | .0457 | .0458 | .0452 | .0410 |
| ICC | 48.90% |  |  |  |  |  |  |
| Change in variance Lev. 2 |  | -1.00% | -18.14% | -14.85% | -11.46% | 7.10% | -29.46% |
| Change in variance Lev. 1 |  | -7.53% | -0.92% | -3.31% | -3.08% | -4.41% | -13.25% |
| N | 64 | 64 | 64 | 64 | 64 | 64 | 64 |
| Groups | 26 | 26 | 26 | 26 | 26 | 26 | 26 |

Notes. *p < .05, ** p < .01. ICC = Intraclass correlation coefficient.

**Table 3.** Models used to predict customer satisfaction.

Multilevel modeling has several advantages over other techniques and has been used frequently for repeated measures data (Hoffman & Rovine, 2007; Quené & Van Den Bergh, 2004). Multilevel modeling is becoming increasingly popular in disciplines such as sociology or behavioral research (e.g. Carvajal et al., 2001; Raudenbush & Bryk, 2002). In our models, measurements occasions are nested within clients groups; thus, level-1 units consist of the repeated measures for each group, and level-2 units are represented by the group. Due to missing data, we were not able to collect the complete email network of the 26 groups over all the 4 data points. However, missing observations are handled in multilevel models, without causing additional complications.

The intraclass correlation coefficient (ICC) is 48.90%, indicating that almost half of the sample variance is attributable to differences in workgroups assigned to customers (level 2). We only included fixed effects into the models for the significant predictors. Random effects of all predictors have been tested by likelihood-ratio tests; however they do not improve the models. Predictors are all centered on the sample mean except for time, thus estimating NPS. Time has been centered on value 1 (as time is coded as 1-2-4-5), thus the intercept in Model 6 estimates NPS at time 1 (i.e. the first wave).

Models are first presented to assess the contribution of each single predictor, and then all significant predictors are included in Model 7. The models show the significant effects of complexity, group betweenness oscillation and average response time. Effect size can be inferred from variance reductions: the variance of the model with predictors is compared with variance of the empty model (i.e. Model 1).

## 5. Discussion

Even though feedback sessions have been demonstrated to be such a critical part of organizational life, there is still a lack of clear understanding of the precise effect that feedback has on productivity and performance (Clampitt & Downs, 1993; Kluger & DeNisi, 1996). This study offers empirical evidence to suggest that promoting self-awareness of communication styles might prompt behaviors that can increase customer satisfaction. Providing employees with a monthly description of their communication patterns is a method that fits with traditional techniques of self-evaluation (Johnston, 1967). Through a process of open dialog, employees are offered a unique opportunity to constantly discuss group dynamics and leadership behavior that are usually taken for granted. This process is essential to nurture the creation of communities where clients and employees participate in a process of knowledge co-creation. The employees who are more committed to mirror their own performance could be nurtured as leading practitioners of communication patterns, becoming lead users who feed lessons learned back to other users. Promoting organizational self-awareness is also key to improve the organization's sensemaking ability, as it helps individuals build a better understanding of events that are novel, ambiguous, confusing, or in some other way violate expectations (Maitlis & Christianson, 2014, p.57). Adopting an organizational sensemaking perspective, we see how people tend to cope with confusing or complex events by sharing perceptions with others and creating meaning through discussion and dialogue (Holt & Cornelissen, 2014).

Our results support the hypothesis that exposing members of the organization to their own communication patterns - triggering their awareness and creating the basis for self-evaluation and behavior change - will increase customer satisfaction (H1). In particular, using simpler language seems to be associated with better customer satisfaction (H3). This confirms the results of studies in the area of service marketing showing that communication effectiveness requires the formal and informal sharing of frequent, significant, clear and timely information between a client and a provider (Morgan & Hunt, 1994; Webster & Sundaram, 2009). H2 is only partially supported since group degree centrality correlates with customer satisfaction, but this variable is not significant in the multilevel models. Customers are likely to be more satisfied if employees answer e-mail faster (H4), thus showing a higher level of commitment. In addition, our findings support the idea that more stable communication patterns, with less oscillations in betweenness centrality, are more conducive to higher customer experience (H5). Figure 3 presents a summary of our results.

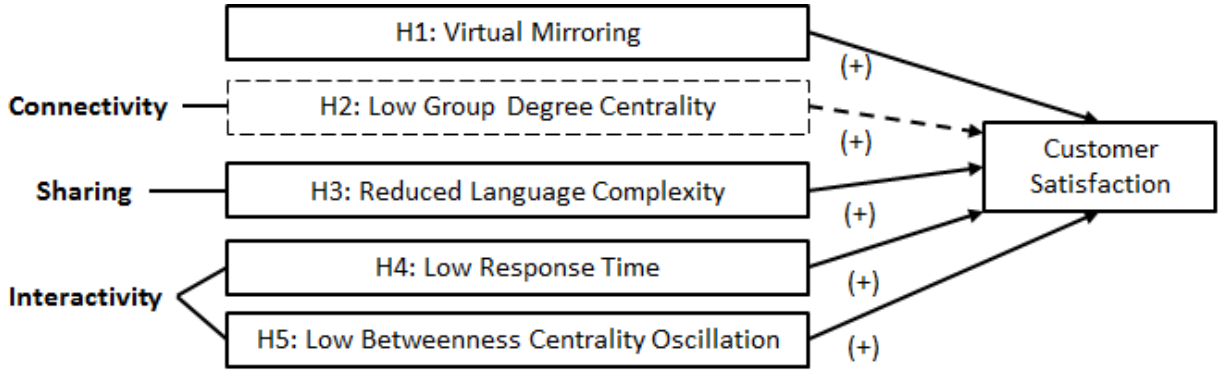

**Figure 3.** Summary of Results (dotted line represents partially supported hypothesis)

This study offers new insights on the measurement of customer satisfaction, which is widely considered an antecedent of customer loyalty, retention and firm profitability (Reichheld, 2003; Wirtz & Lee, 2003; Bearden & Teel, 1983). The results demonstrate not only the importance of relational concepts such as involvement or commitment to the customer, but also a deeper level of engagement, which has been recognized in marketing literature as a strategic imperative for generating enhanced performance (Brodie et al., 2011).

Our findings function as reminder for managers of service-based companies that interpersonal communication has a critical role to play in influencing the customers' total experience and satisfaction. Customers forms specific perceptions of service quality based on the nature, frequency and effectiveness of communication (Patterson and Patterson, 2016).

The approach we propose in this paper overcomes some of the limitations of the survey-based approaches. The same tool we used to collect customer satisfaction data, NPS, has been criticized for being attitudinal rather than behavioral, measuring how many people say they would be likely to recommend a product or service, rather than how many are actually doing so (Keiningham et al., 2007). In this paper, we suggest using new metrics such as complexity of subject lines and rotating leadership to overcome some of the limitations of traditional tools. Instead of relying exclusively on traditional methods of gaining customer feedback, such as phone or mail surveys, interviews or focus groups, it is now possible to also use computer programs to filter out the relevant data and computationally treat opinions and subjectivity in text.

## 6. Research Limitations, Future research and Conclusions

While mirroring sessions represent a less expensive and complementary method to promote organizational awareness, we need to recognize their potential costs, which are largely related to potential psychological effects on participants. This is a common problem of many feedback interventions, which are all based on the assumption that a positive feedback equates with reinforcement and a negative one equates with punishment (Kluger & DeNisi, 1996). We believe our experimental results offer an important contribution to the extant literature trying to shed light on the effect of feedback interventions in organizations. Creating an organizational structure that helps employees reflect on their level of engagement with customers via their own communication will help them make sense of the situation around them, reduce ambiguity, and establish meaningful relationships within and across organizational boundaries (Holt & Cornelissen, 2014). By correlating communication patterns with customer satisfaction we find that variables such as responsiveness and steady communication are important factors to predict customer satisfaction. By mirroring these factors back to the employees, we can start a process of self-awareness and personal development that will have a positive impact on customer satisfaction.

This study is based on an experiment run at a large global service company. The nature of the industry represents a possible limitation on how we interpret data. In the service industry there is a demand on employees to be more sensitive to customer needs because that is the measure of effectiveness, while manufacturing employees might use other measures of effectiveness that focus less on personal relationships (Clampitt & Downs, 1993). However, workers in different industries rely more and more on digital communications to get their job done, thus we would not be surprised to obtain similar results in other knowledge-intensive industries. Our study focused mainly on managers' communication behaviors, thus our results might not be generalized to employees at all levels.

Another limitation is represented by the lack of access to email content, since we were constrained by privacy reasons to use the e-mail subject lines. Since we have a large dataset, spanning multiple years, with millions of e-mail messages, we were still able to mine millions of subject lines. We recognize this as a potential limitation, therefore we used the subject line only to calculate language complexity rather than calculating more sophisticated indicators such as sentiment and emotionality (Gloor, 2016). Future research should include the calculation of emotionality and complexity of the language used by mining the body of the email. An increasing number of studies have used the email body to assess change management initiatives and team development (Zhang et al. 2013; Grippa et al., 2012). For example, Zhang et al. (2013) calculated positive and negative sentiment in the e-mail text using a simple "bag-of-words" approach, employing a user-generated mood word list, relevant for the domain area to calculate positivity and negativity of an e-mail's content. In addition, future research should control for other variables such as employees' personality traits, work schedules (to test the influence on the average response times), level of friendliness or listening skills - often cited as predictors of higher customer satisfaction (Motley, 2003; Gremler & Gwinner, 2000). Unfortunately, the privacy and confidentiality policies at the organization where we run the experiment did not allow for any other data gathering. In addition, we suggest replicating our experiment in different business contexts and possibly with similar starting NPS values for the experimental and the control groups.

It is important to recognize the ethical issues that any organizational network analysis typically involves, especially when tools are now able to mine a large amount of data. When showing results to management, it is critical to help them interpret the results in a way that does not penalize any other organizational member. We do not encourage managers to target specific individuals whose communication behaviors show a different trend in centrality or responsiveness, as there might be reasons beyond the control of the individual (for example being ill, or being in vacation). The intent of our study was to educate organizational members to recognize their own communication patterns and reflect on their position in the communication networks. To protect privacy of individual respondents, confidentiality and anonymity should be clearly embedded into any social network interventions (Borgatti & Molina, 2005).

**Appendix**

**Workflow for the Virtual Mirroring process (Figure 4)**

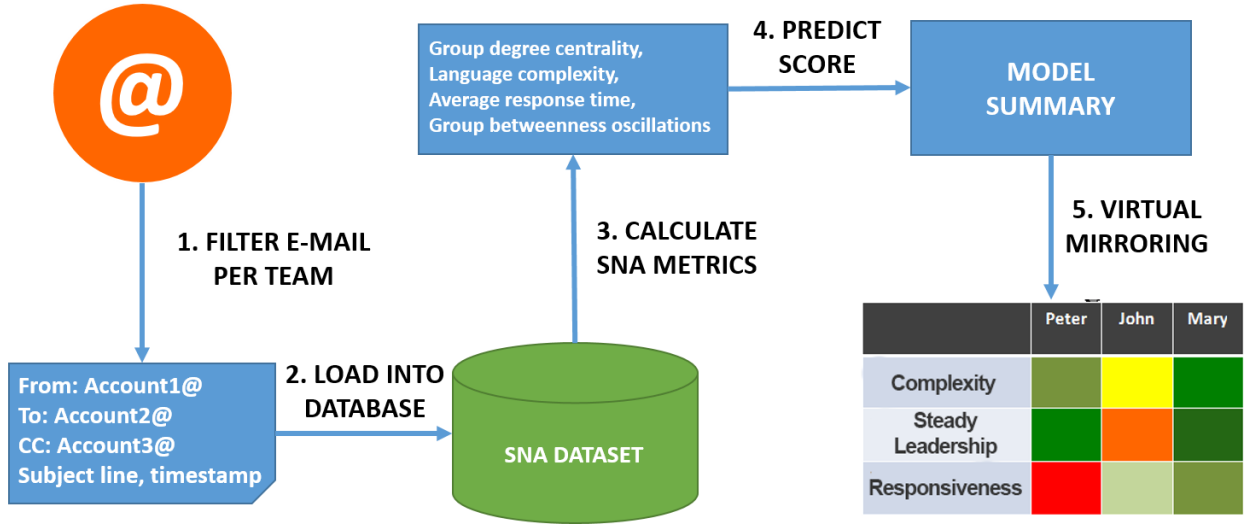

**Figure 4.** Workflow for the Virtual Mirroring process.

**Collaboration Scorecard shown to account leaders (Table 4)**

Sample scorecard mentioned in (Gloor & Giacomelli 2014). Note that we used a part of this scorecard (Complexity, Steady Leadership and Responsiveness). This because some of the variables changed in our experiment: as account leaders changed their behavior, new variables gained predictive power, while others lost it.

| Metric | Well-performing | Low performing | Remedy | SNA Metric |
|---|---|---|---|---|
| Central leaders | Leaders are well connected to customers indicating recognition by the customer | Leaders are not recognizable from the outside | Designated leaders need to become more pro-active | Betweenness centrality Degree centrality |
| Steady leadership | The same leaders stay in permanent contact with the client, providing a consistent contact point | There is continuous rotation among the central leaders, making it hard for clients to identify reliable touch points | A few people need to be designated as customer contacts, and communication has to be channeled through them | Oscillation in group/actor betweenness centrality |
| Responsiveness | Prompt and consistent response indicates high team and customer engagement; enables fast problem-solving | Slow response time might indicate a disengaged team, which often correlates with a dissatisfied customer | Designated customer contacts should answer faster; raise awareness of importance of speed | Average Response Time (ART) |
| Initiative | Proactive communication between parties; initiative is frequently on the leaders' side. A few key leaders send more than they receive | Uncoordinated flood of e-mails from disparate sources leads to confusion, and swamps client e-mails | Customer communication should be channeled through designated touch points | Average Weighted Variance in Contribution Index (AWCI) |
| Sentiment | Consistently well-balanced sentiment of email subject lines signals fair and fact-based discussion | Uneven or overly positive and overly negative sentiment of email subject line may indicate severe customer dissatisfaction | Don't use overly positive or overly negative language, but stick to the facts. | Calculated sentiment based on terms' positivity and negativity |
| Complexity | A simpler and more shared language is used in the communication between parties | A more complex language is used, making the message more difficult to understand | Use a shared language that the customer can understand | Language Complexity |

**Table 4.** Collaboration Scorecard.